\documentclass[apj]{emulateapj}

\usepackage{graphicx}
\usepackage{microtype}
\usepackage[]{hyperref}  % PDF hyperlinks, with coloured links
\def\link_col{blue}
\hypersetup{colorlinks,citecolor=\link_col,filecolor=\link_col,linkcolor=\link_col,urlcolor=\link_col}

\slugcomment{To appear in The Astrophysical Journal Letters}

\shorttitle{Magnetized Fermi Bubble walls}
\shortauthors{Jones et~al.}

\begin{document}

\title{Magnetic substructure in the northern Fermi Bubble revealed by polarized microwave emission}
\author{David I. Jones\altaffilmark{1}, Roland M. Crocker\altaffilmark{1}, Wolfgang Reich\altaffilmark{2}, J\"{u}rgen Ott\altaffilmark{3} \& Felix A. Aharonian\altaffilmark{4,1} }
\altaffiltext{1}{Max-Planck-Institut f\"{u}r Kernphysik, Postfach 103980, 69029 Heidelberg, Germany.}
\altaffiltext{2}{Max-Planck-Institut f\"{u}r Radioastronomie, Auf dem H\"{u}gel 69, 53121 Bonn, Germany}
\altaffiltext{3}{National Radio Astronomy Observatory, PO Box O, 1003 Lopezville Road, Socorro, NM 87801, USA}
\altaffiltext{4}{Centre for Cosmic Physics, Dublin Institute for Advanced Study, Dublin D4, Ireland.}

\email{djones@mpi-hd.mpg.de}

\begin{abstract}
We report a correspondence between giant, polarized microwave structures emerging north from the Galactic plane near the Galactic center and a number of GeV gamma-ray features, including the eastern edge of the recently-discovered northern Fermi Bubble. 
The polarized microwave features also correspond to structures seen in the all-sky 408~MHz total intensity data, including the Galactic center spur. 
The magnetic field structure revealed by the WMAP polarization data at 23~GHz suggests that neither the emission coincident with the Bubble edge nor the Galactic center spur are likely to be features of the local interstellar medium.
On the basis of the observed morphological correspondences,  similar inferred spectra, and the similar energetics of all sources, we suggest a direct connection between the Galactic center spur and the northern Fermi Bubble.

\end{abstract}

\keywords{Galaxy: center --- Galaxy: kinematics and dynamics --- Galaxy: structure --- ISM: jets and outflows --- ISM: magnetic fields --- ISM: structure}

\section{Introduction}\label{sec:intro}
Recently, spectacular evidence has emerged for enormous features in the $\sim$ GeV gamma-ray sky observed by the $Fermi$-LAT instrument \citep{Michelson2010}: bilateral lobes or `Bubbles' of emission centered on the core of the Galaxy and extending to $\pm50^\circ$ ($\pm10$~kpc assuming a Galactic center (GC) origin) from the Galactic plane \citep{Su2010}. 
The presence of large-scale X-ray emission found in the ROSAT data \citep{Snowden1997,BlandHawthorn2003} and the WMAP `haze' (at lower Galactic latitudes; \citealt{Finkbeiner2004,Dobler2011a}) helps support the notion that the Fermi Bubbles constitute evidence for an anomalous population of high-energy, hard-spectrum, non-thermal particles pervading a large portion of the Galactic halo.
The foremost question about the Bubbles has now become where these particles originate.
Due to the large luminosity of $\sim4\times10^{37}$~erg~s$^{-1}$ in the gamma-ray domain (an order of magnitude larger than the Bubbles' microwave luminosity but more than order of magnitude less than their X-ray luminosity; \citealt{Su2010}); a hard spectrum of $dN/dE\sim E^{-2}$ from 1--100~GeV; their vast extent;  and their relatively uniform  gamma-ray intensity, the Bubbles are difficult to explain in a consistent manner.

The Bubbles' emission has been hypothesized \citep{Dobler2010} to be produced by the same population of highly-relativistic cosmic-ray electrons which emit synchrotron radiation at tens-of-GHz frequencies and are posited to simultaneously produce $\gtrsim1$~GeV gamma-rays through the inverse Compton process.
Given the severe radiative energy losses that such electrons would experience, however, they would either have to be transported at $\gtrsim$3\% of $c$ from the GC or accelerated throughout the volume of the Bubbles \citep{Crocker2011}.
The former presumably implies  an AGN-like outburst  from the central black hole, Sgr~A*, in the past few million years \citep{Su2010,Guo2011,Zubovas2011}; the latter might be explained as due to all-pervading shocks \citep{Cheng2011} or distributed, stochastic, acceleration on plasma wave turbulence \citep{Mertsch2011}.
Alternatively, \citet{Crocker2011} have proposed that a cosmic-ray $proton$ (and heavier ion) population -- associated with extremely long timescale star formation in the GC and carried out of the plane on a wind -- can naturally explain the Bubbles'  gamma-ray emission provided the protons are trapped for timescales approaching $10^{10}$~years. 
The idea that the Bubble emission may arise from the decay or annihilation of dark matter particles \citep[e.g.,][]{McQuinn2011} would now seem to be disfavored on morphological grounds \citep{Su2010} \citep[though see][for an alternative point of view]{Dobler2011}.

One important test for any putative explanation of the Bubbles is to account for their reported sharp edges. 
The protonic scenario, in particular, would seem to require some sort of magnetic structure  able to confine the Bubbles' non-thermal proton population for long times.
Such `magnetic walls' could be revealed as thin, elongated structures at radio and microwave frequencies
due to synchrotron emission from either primary or secondary electrons (the latter being naturally predicted in the protonic scenario).
If, in addition, the foreground Faraday screen were not too large and the magnetic field topology  sufficiently regular, such emission may exhibit  detectable polarization.
 
While an indirect search for polarized emission coincident with the WMAP haze by \cite{Gold2010} has yielded only an upper limit this is reconcilable with this microwave emission being due to synchrotron if the magnetic field structure is highly tangled \citep{McQuinn2011}. 
In addition \citet{Dobler2011a} has recently claimed that the noise in the polarization data is likely too large to allow for the detection of hard-spectrum emission (such as that expected from the WMAP haze in total intensity data).
Considering the above arguments, along with both the claim of \citet{Su2010} that the X-ray emission is limb-brightened on size-scales of $\sim12'$ (across the edge), and the fact that the emission in the northwestern part of the lobe appears brighter -- at least towards the Galactic plane -- together suggest that polarized emission might be detected on these scales.

Indeed, in this $Letter$ we show that  the 5-year WMAP polarization data reveals  magnetic structures coincident with i) part of the eastern edge of the gamma-ray detected northern Fermi Bubble reported by \citet{Su2010} and ii) the so-called GC spur discovered in unsharp-masked\footnote{Unsharp-masking is a method for obtaining separate small- and large-scale emission images using Gaussian convolution -- in contrast to Fourier transformational methods -- so as to provide images with as high a fidelity as possible; see \citep{Sofue1979} for more information} 408~MHz data \citet{Sofue1989}.

\begin{figure}[t]
\centering
\includegraphics[width=0.45\textwidth]{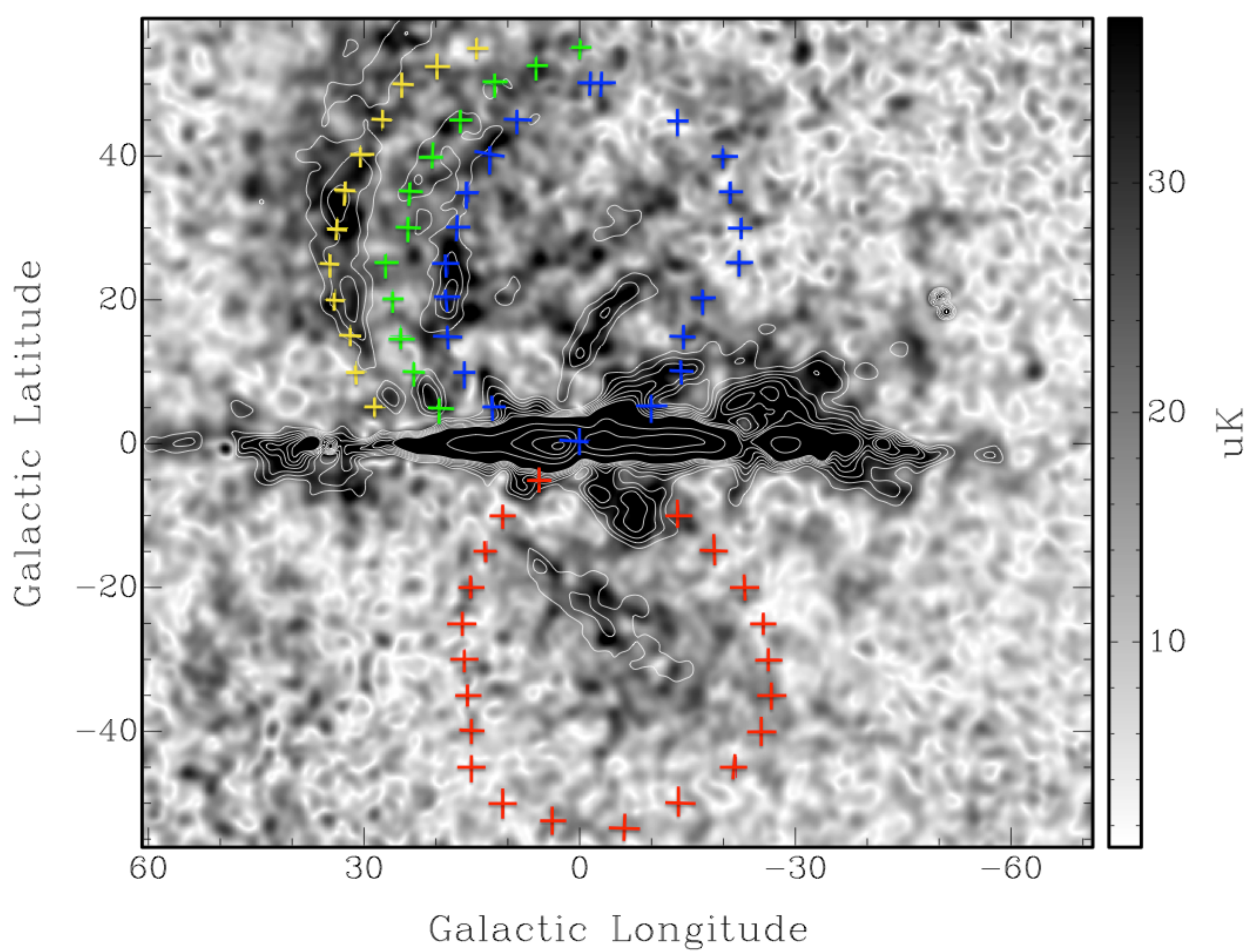}
\caption{View of the inner Galaxy in 33~GHz (Ka-band) total polarized intensity overlaid upon a 23~GHz (K-band) 5-year WMAP data in total polarized intensity (white) contours at 70, 88, 105, 123, 140, 158, 175, 350, 530, 700, 878, 1054, 1230, 1400, and 1580~$\mu$K. Both data-sets have been convolved with a Gaussian beam of FWHM $2^\circ$. The crosses denote the positions of the northern Bubble edge (blue), inner (green) and outer (yellow) northern arcs and southern (red) Bubble edge, respectively, and taken from Table~1 of \cite{Su2010}. The emission in the southern Bubble is taken to be a continuation of the North Polar Spur, and thus not of consideration here.
\label{fig:BubblePlot}}
\end{figure} 
 
 \section{WMAP data processing}\label{sec:data}
\begin{figure*}[t]
\centering
\includegraphics[width=0.7\textwidth]{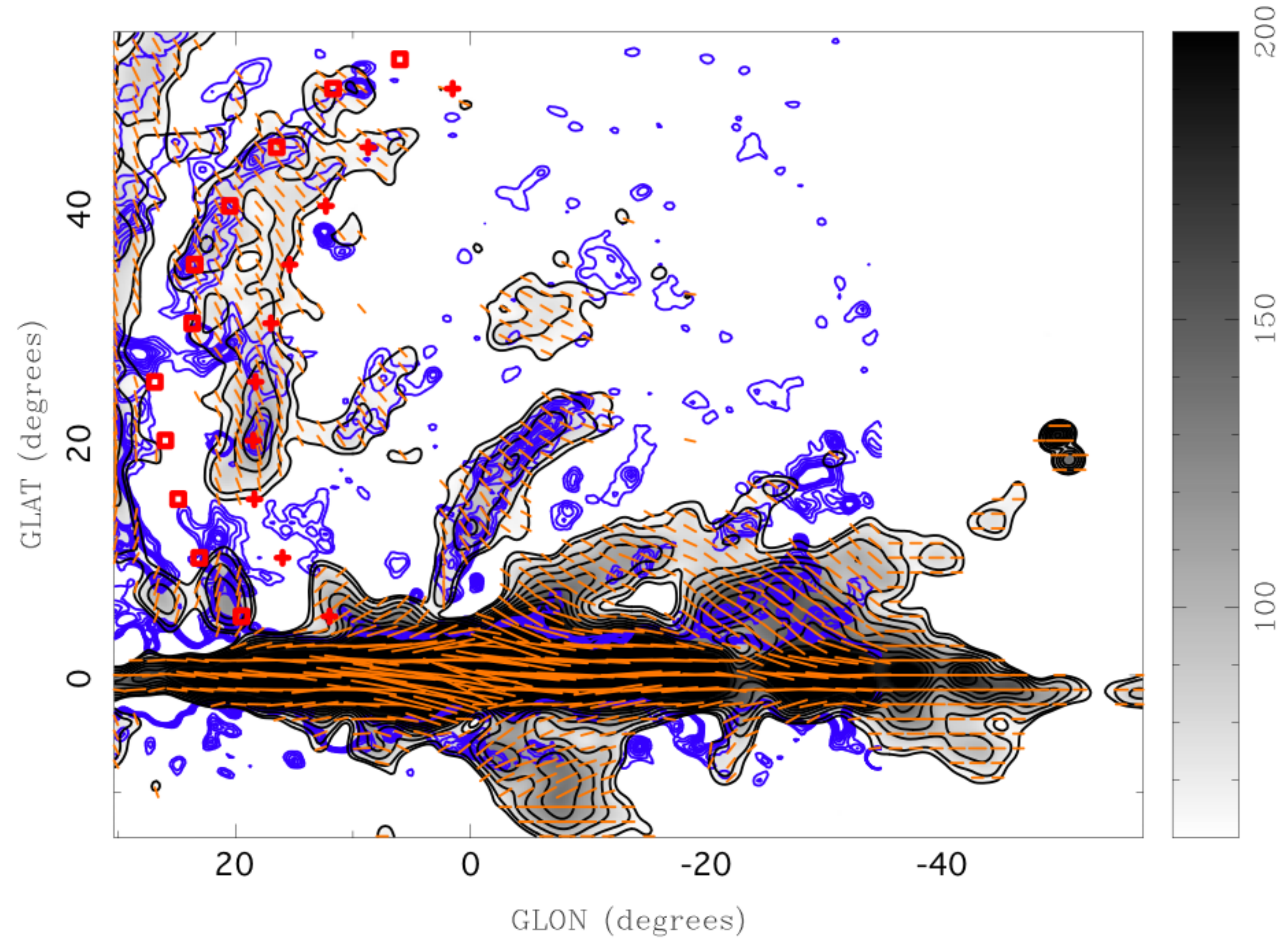}
\caption{
Image of the northern Bubble section comprised of WMAP 23~GHz (K-band) polarized intensity overlaid with the same (black) contours as in fig. \ref{fig:BubblePlot}, and (blue) 408~MHz contours at levels of 6, 8, 10 and 12~K. Also plotted are vectors with length proportional to the polarized intensity and orientation at 90 degrees to the observed polarization direction  (i.e., tracing the magnetic field orientation).  The 408~MHz image has had the large-scale emission filtered above $2.0^\circ$ (i.e., the size of the Gaussian beam of the large-scale emission image). The intensity scale runs from 60 to 200~$\mu$K as indicated by the colored wedge at the right. The Fermi Bubble edges are indicated by the crosses, whilst the inner and outer northern arcs are denoted by squares and circles respectively.  
\label{fig:WMAP408MHz}
}
\end{figure*}

We obtained the 5-year WMAP data products as a function of frequency (23, 33, 41, 61 and 94~GHz, corresponding to the K, Ka, Q, V and W bands respectively) from the data server, LAMBDA\footnote{\url{http://lambda.gsfc.nasa.gov/product/map/current/}. We also note here that although the 7-year data is available, there should be negligible differences.}. 
The images obtained from this server are contained in nested, HEALpix\footnote{HEALpix software and documentation can be found at \url{http://healpix.jpl.nasa.gov}.} FITS file format which is natively in Mollweide projection. The microwave data were rendered into the native projection of the 408 and 1420~MHz data used in figure~\ref{fig:WMAP408MHz} and section \ref{sec:equipartition} to the same projection.
These images contain the Stokes I, Q and U in a single image, so we exported the individual Stokes images into stand-alone FITS images for each polarization and band. 
 We then used the MIRIAD data reduction software package\footnote{available at \url{http://www.atnf.csiro.au/computing/software/miriad/}} to export the FITS files into the MIRIAD format and then placed the correct intensity units (mK of brightness temperature) and requisite beam size into the header (these values were taken from Table 5 of \cite{Page2003}, and although the WMAP beam is not a Gaussian, the beam-size cancels out in any flux density integration). 
We then subtracted the internal linear combination (ILC) image from the Stokes I images at every frequency. The ILC images are the foreground-subtracted images containing only cosmic microwave background (CMB) signal, and thus we subtract this (albeit non-dominant) image, so that we do not confuse these structures with our signal

Such images (the ILC images) are not supplied for the Stokes Q and U data (and hence are not subtracted from the polarized intensity images). This is because the expected polarized intensity of the CMB signal is dominated by the Galactic synchrotron (at low frequencies -- i.e., 23~GHz) and dust (at high frequencies -- i.e., 94~GHz; see \citet{Page2007} for more information).
In producing the polarization images, we do not subtract any of the background images that the WMAP consortium produces. This is because we are searching for a Galactic signal that is most likely due to synchrotron emission.  As \citet{Page2007} has shown, polarized emission due to (thermal) dust is not expected to dominate over other emission processes until at least 90~GHz.

From the Stokes Q and U data at each waveband, we then created polarized intensity and polarization angle images of the WMAP data for each frequency band by taking each Stokes Q and U images from each waveband and combining them using the standard equations for polarized intensity and angle.
In the regions that we sample, we find an average 1$\sigma$ r.m.s. noise of 50~$\mu$K for the K, and Ka bands (23 and 33~GHz) for the polarized intensity images.
We note here that although we produced images of polarized emission at 41, 61 and 94~GHz, they do not show the same structures that the 23 and 33~GHz images.

\section{Results}\label{sec:results}
Figure \ref{fig:BubblePlot} shows an image of the 33~GHz (Ka-band) polarized intensity emission, overlaid with contours of the 23~GHz WMAP polarized intensity. 
Also shown in figure~\ref{fig:BubblePlot} are the co-ordinates of various features identified in the $Fermi$-LAT gamma-ray data by \citet{Su2010} including the structures labeled the inner and outer northern arc and the edges of the northern Bubble.
There is a clear morphological coincidence between part of the eastern edge of the northern Bubble and a region of enhanced polarized microwave emission.

Coincident, enhanced emission from the same eastern edge region is also visible in an unsharp-masked \citep{Sofue1989}, total-intensity rendering of the all-sky 408 MHz data from \citet{Haslam1982} (cf. figure~\ref{fig:WMAP408MHz}).
This same figure reveals other regions bright in both 408 MHz total intensity and polarized microwave intensity like the North Polar Spur, which is thought to be a local interstellar medium feature (note that there seems to be some confusion between the outer northern arc and the North Polar Spur).
Of particular interest, however, is a structure emerging from directly north of the Galactic centre, extending to $\sim25$ degrees Galactic latitude and appearing to bend by $\sim15^\circ$ to Galactic west above $\sim10$ degrees latitude, assuming a GC origin.
This hard-spectrum ($\gamma\sim-2.2$), non-thermal feature is the GC Spur previously discovered in the large-scale filtered, unsharp masked 408 and 1420~MHz radio continuum data \citep{Sofue1989}.
Interestingly, a gamma-ray spur, roughly coincident with the GC Spur has previously been reported in EGRET data -- although the statistical significance of this feature could not be determined \citep{Hartmann1997}.
The GC Spur also appears partially coincident with  linear substructure recently detected within the northern Bubble\footnote{Douglas Finkbeiner, {\it The Emerging, Multi-Wavelength View of the Galactic centre Environment}, Heidelberg, Germany, October 2011}.
We cannot positively identify the same structures in the higher-frequency microwave maps above the noise limits of 74, 105, and 219~$\mu$K at 41, 61 or 94~GHz respectively, consistent with a falling, non-thermal spectrum.
We also note here several other interesting features. We find diffuse emission within the confines of the southern Bubble (c.f. figure \ref{fig:BubblePlot}), which is taken to be associated with the North Polar Spur, and thus not associated with the southern Bubble.
More interesting, however, the magnetic field vectors in figure \ref{fig:WMAP408MHz} reveal a feature which seemingly runs from Galactic north-west -- from $(l,b)\sim(-15^\circ,\sim10^\circ$) -- through the GC to the south-east -- to $(l,b)\sim(10^\circ,-5^\circ)$ -- and is aligned with similar features found in the linearly polarized microwave and the 408~MHz radio continuum emission.
This feature is also aligned with the (blue and red) crosses of the outer limits of the northern and southern Bubble walls as illustrated by figure \ref{fig:BubblePlot}.

Finally, we note that although we have observed polarized microwave emission coincident with the eastern edge of the northern Bubble, this correspondence is not perfect; we do not observe such a correspondence from other parts of the Bubbles. 
We do not preclude, however, that deeper observations would reveal more of the Bubbles' edges.

\section{Discussion}\label{sec:discussion}
\subsection{The Galactic Center Spur and the Fermi Bubbles}
There are a number of pieces of circumstantial evidence relating the Fermi Bubbles to the GC Spur.
Firstly, the GC Spur's extension is entirely contained within the 50 degree height of the northern Fermi Bubble, with both appearing to emanate from the Galactic plane at the position of the GC.
Secondly, both are characterized by $\gamma\sim-2.2$ spectra indicating emission from a hard, non-thermal particle population.
Finally, the energetics are a sensible match: the gamma-ray luminosity of the northern Bubble is $\sim 2 \times 10^{37}$ erg/s; given the covering fraction of the GC spur over the Bubble, the estimated (\citealt{Sofue1989}) $\sim 10^{36}$ erg/s synchrotron luminosity of the former is reasonable if the structures are related.

Additional supporting evidence can be gleaned by observing that the GC Spur is curved noticeably to Galactic west. 
This morphology is reminiscent of the magnetic field structure recently determined for the star-formation-driven outflow from NGC253 in which field lines, though vertical close to the disk, become noticeably `wound' up by differential rotation with increasing height \citep{Heesen2011}.
If we were to associate the GC Spur with a GC star-formation event occurring in the $r\simeq100$~pc star-forming ring discovered by \citet{Molinari2011} rotating at $\sim$80 km/s, this would imply a rough timescale of $\pi \ r/ v_{rot} \simeq 3$~Myr (note that the structure has not gone through a full turn).
Assuming, then, this timescale and adopting a total height of the GC Spur of $\sim$ 4 kpc, we find that the outflow should be rising at a speed $\sim$ 1200 km/s.
This value is towards the upper end of outflow speeds allowable in the analysis of \citet{Crocker2011b}, thus strengthening the connection between the GC Spur and the outflow implied (using completely independent arguments) in \citet{Crocker2011b}.

\subsection{Magnetic structures of the GC Spur and Fermi Bubble edges}
Figure \ref{fig:WMAP408MHz} also shows the position angle of the polarized emission, rotated by $90^\circ$, to indicate the inferred orientation of the magnetic field.  
Faraday rotation at these high frequencies is not important out of the Galactic plane.
The magnetic field lines in the northern Bubble's eastern wall and the GC Spur both run roughly perpendicular to their extension, consistent with a toroidal field structure and suggesting that neither of  these features is related to a local supernova remnant shell (in which one would expect a predominantly tangential field; cf. the field lines in the North Polar Spur). 
In support of the claim that the GC Spur is {\it not} a feature of the local interstellar medium, we note that this structure is almost indiscernible in 1.4~GHz polarized intensity data \citep{Testori2008,Wolleben2006}, in contrast to the microwave data, which is consistent with the magnetic horizon effect \citep{Landecker2002} and this structure being located at a distance larger than a few kpc.

\subsection{Estimation of the equipartition magnetic field amplitude of the Bubble walls}\label{sec:equipartition}
We have, via recourse to the 408~MHz all-sky data of \citet{Haslam1982} and a combination of the Stockert 25-m data of \citet{Reich1982} and \citet{Reich1986} and Villa Elisa 30-m data of \citet{Reich2001}, estimated the equipartition magnetic field contained within these structures. 
To do this, we integrated the temperature brightness within the lowest (80~mK) 23~GHz contour of figure \ref{fig:BubblePlot}, obtaining $\sim2400$ and $\sim1100$~Jy at 408 and 1420~MHz respectively. 
This gives a spectral index of $\alpha\sim-0.63$ (for $S_\nu\propto\nu^\alpha$), suggesting that the emission is predominantly  synchrotron, whilst a radio flux of $\sim2400$~Jy at 408~MHz gives a radio power of $\sim2\times10^{19}$~W~Hz$^{-1}$ assuming a distance to the structure of 8~kpc.
Assuming a pure power-law distribution of the emitting electrons, 
integrating the emission spectrum from 10~MHz to 10~GHz gives a total luminosity of this structure on the order of $\sim 6 \times 10^{35}$~erg~s$^{-1}$.
This is similar to the luminosity of the GC Spur, and reinforces our arguments above that these structures are of a GC origin and that they fit well, energetically, with the larger Bubble structures.

Utilizing the above-stated spectral index and synchrotron luminosity, we estimate an equipartition magnetic field value for the Bubble wall, using equation 3 from \citet{Beck2005}, of $B_{eq} \simeq 15 $~$\mu$G $[({\mathbf K}_0/100)(l/\textrm{kpc})^{-1}f^{-1} (\Omega/0.055 \ \textrm{sr})^{-1}]^{0.28}$, where we have normalized to 
a high-energy proton to electron number density ratio of {\bf K}$_0=100$, a path length through the emitting medium $l = 1$ kpc, a filling factor of $f = 1$, and a total solid angle $\Omega \sim 30^\circ\times6^\circ = 0.055$ sr.
We note that -- given the structure is likely far from the plane -- {\bf K}$_0$ might be significantly larger than 100 (because of the much more severe energy losses faced by electrons relative to protons).

The structures shown in figure \ref{fig:WMAP408MHz} at 408~MHz (and also at 1420~MHz) correspond well to the polarized microwave structures that we argued above are the same structures identified in \cite{Su2010} as the northeastern Fermi Bubble walls, inner and outer northern arcs.
The spectrum of these objects at low-frequency in radio continuum emission is somewhat harder than the $\sim 0.7$ due to the general, disk cosmic-ray electron population.
The inferred spectrum of the synchrotron-emitting parent electron population, $2 \alpha + 1 = 2.26$, is also hard and consistent within uncertainties with the hard spectrum of the non-thermal particle population responsible for the Bubble gamma-ray emission.
Based on their spectrum and the energetics we argue that these structures are the same as those observed in polarized microwave and gamma-ray emission.

\section{Conclusions}\label{sec:conclusions}
We have presented images of the polarized and total intensity at 23~GHz (for the polarized emission) from WMAP data and 408~MHz (for unsharp-masked, total intensity emission)  revealing structures which emerge north from the Galactic plane in the vicinity of the GC coincident with structures observed in  $Fermi$-LAT gamma-ray data. 
The positional coincidence of these structures and other pieces of evidence cited above suggest that these are the same or related structures seen at different wavelengths.
The magnetic structures of both the eastern wall of the northern Bubble and the GC Spur inferred from  the polarized
23~GHz data suggest that neither of these structures in likely to be a local supernova remnant shell.

Finally, we suggest that these objects deserve much (observational) scrutiny, since they may constitute evidence that the central part of our Galaxy is -- and/or once was -- more active than previously thought. 
A dedicated observational program, using the new generation of wide-band radio telescopes, could deduce the location of these objects (through polarization studies) and give more clues as to the origin of the Fermi Bubbles.

\acknowledgments The authors would like to thank Rainer Beck, Doug~Finkbeiner,~and~Meng~Su~for~illuminating~conversations~and/or~correspondence. RMC would particularly like to thank Tracy~Slatyer for discussions. 
We also thank the anonymous referee for comments which greatly improved the manuscript.

\end{document}